Digital poster 2772

# Flow 2.0 - a flexible, scalable, cross-platform post-processing software for real-time phase contrast sequences

Pan Liu[1,2], Sidy Fall[1], Olivier Baledent[1,2]

[1] CHIMERE UR 7516, Jules Verne University, Amiens, France.

[2] Medical Image Processing Department, Jules Verne University Hospital, Amiens, France.

## Summary of Main Findings (250 characters, 239/250)

Flow 2.0 is an end-to-end easy-of-use software that allows us to quickly, robustly and accurately perform a batch process real-time phase contrast data, and multivariate analysis of the effect of respiration on cerebral fluids circulation.

## Synopsis (99/100)

Real-time phase contrast sequences (RT-PC) has potential value as a scientific and clinical tool in quantifying the effects of respiration on cerebral circulation. To simplify its complicated post-processing process, we developed Flow 2.0 software, which provides a complete post-processing workflow including converting DICOM data, image segmentation, image processing, data extraction, background field correction, anti-aliasing filter, signal processing and analysis and a novel time-domain method for quantifying the effect of respiration on the cerebral circulation. This end-to-end software allows us to quickly, robustly and accurately perform batch process RT-PC and multivariate analysis of the effects of respiration on cerebral circulation.

## Introduction (850 words to conclusion 816/850)

With further development of imaging techniques and instrumentation, real-time phase contrast sequences (RT-PC) have become possible[1]. Compared to conventional phase-contrast sequences, its most significant feature is the quantification of cerebral blood flow or cerebrospinal fluid (CSF) in "real-time". Although not yet applied in clinical practice, recent studies have shown that RT-PC has a great potential for non-invasive quantification of the effects of respiration on cerebral fluids circulation[2-4]. However, the fast imaging speed that impairs RT-PC image quality, and a large number of images make it more difficult to post-process RT-PC data. In general, the post-processing of RT-PC data often requires a lot of manual operations and several software to perform the different steps such as region of interest (ROI) segmentation, background field correction, anti-aliasing correction, extraction of flow signals, and finally quantify the effect of respiration on blood flow by comparing respiration and flow signals. To our knowledge, there is no post-processing software specifically dedicated to RT-PC data. Based on the original version[5], we have developed Flow 2.0 software integrating all the above-mentioned features and allowing a high level of automation for end-to-end quantification of the effect of respiration on cerebral blood flow.

## Methods

Flow 2.0 was developed using IDL (Interactive Data Language) and was compiled into a 150Mb executable installation-free program which is available on Windows, Linux and Mac systems. Figure 1 shows the post-processing workflow diagram of the software, it contains two major post-processing steps, the flow quantification (Figure 2) and the quantification of respiratory effects (Figure 3).

**Flow quantification**

**Data extraction,** can import the RT-PC images series and the parametric data from the DICOMDIR file according to the conformance statements of the different manufacturers. **ROI segmentation,** is consisted of two semi-automatic algorithms. One is based on region growth and active contour algorithms to segment vessels with large deformations or displacements during acquisition such as arteries. A second is based on a fully automatic segmentation algorithm in the frequency domain that can easily identify pixels in which dynamic signals are synchronized with the cardiac frequency. This one is used for ROIs with insignificant displacements such as CSF and venous sinuses. **Background field correction,** is a necessary step to correct eddy current-induced velocity quantization errors[6]. Typically, a ROI must be drawn manually in the stationary tissue. Our software includes a background field correction algorithm that selects automatically the stationary tissues around the selected ROI. It used the characteristics of the phase and amplitude in the pixel to recognize static tissue. **Curves flows,** are calculated and displayed in the viewing interface. The main key-parameters of the flow curve (flow rate, stroke volume, areas of the ROI, maximum velocities, …) can be automatically calculated and presented in the viewing interface as well as with ECG and respiratory signals if recorded. In addition, the software also includes anti-velocity aliasing[5] and image denoising functions.

A frequency domain signal processing tool interface that is useful for other processing functions was also developed. An **Export** function can save all the results in a text file, easily importable by other software.

**Quantification of respiratory effects**

The first step is to separate all the individual Cardiac Cycle Flow Curve (CCFC) of the total flow calculated in the ROI and across the entire images of series. This **Signal segmentation**, automatically finds the maximum flows systolic points, uses them to define the period and allows to cut the continuous flow rate signal into multiple independent CCFCs. **Signal reconstruction,** the CCFCs are either labelled as expiratory interval or inspiratory interval based on the recorded respiratory signal. Two average expiratory and inspiratory CCFCs are calculated and used to compare the two respiratory periods. *$Diff_{Ex-In}$ (parameter, delay) signal*, allows to compare each parameter of CCFC (amplitude, mean flow, stroke volume, cardiac period) between the expiratory and the inspiratory period. We hypothesis that breathing impacts blood and CSF flows with a possible delay between the flow curves and the signal recorded by the respiratory belt sensor. This delay was calculated by applying a phase shift on the respiratory signal interval that produce the maximum $Diff_{Ex-In}$ for the considered parameter of interest. **Extraction results,** uses the maximum value of $Diff_{Ex-In}$ *(parameter, delay)* to indicate the intensity and *delay* of the effect of respiration on this *parameter* of blood or CSF flow.

## Results

Figure 4 shows an example of an in vivo application of the software. The automatic algorithm accurately segmented the CSF Roi in the spinal canal. We can see the flow rate signal and the respiratory quantification results.

## Discussion & Conclusion

We have developed a post-processing software for RT-PC that integrates the required functions, with a high degree of automation, with low requirements for a priori knowledge of anatomy, high robustness and supporting functional extensions. The software is compatible with the DICOM files generated by the MRI vendors. Our development makes the post-processing of RT-PC substantially easier, while also providing a multivariate approach to analyze the effects of respiration on cerebral fluids circulation.

Digital poster 2772

Figures (4/5)

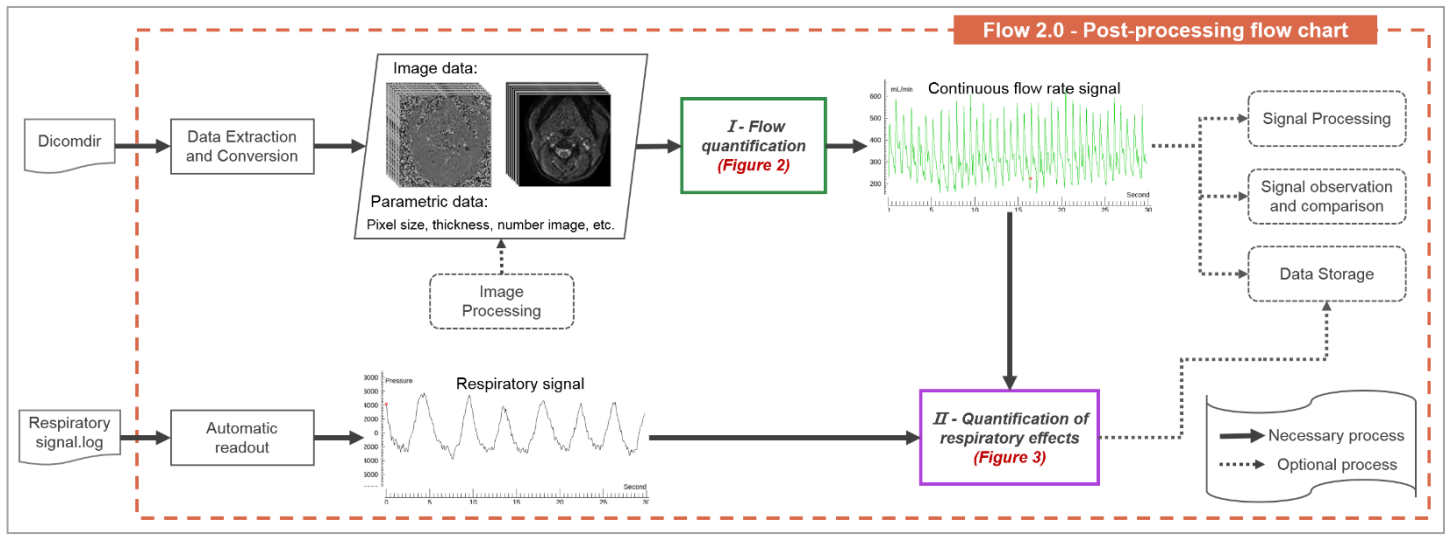

Figure 1: **RT-PC post-processing workflow diagram of Flow 2.0.** The RT-PC sequence images can be automatically extracted by reading the DICOMDIR file, and the flow rate signal can be extracted by the flow quantification processing phase (green box & figure2). The flow rate signal and the respiratory signal can be used as input parameters to quantify the effects of respiration on the flow signal (purple box & figure 3).

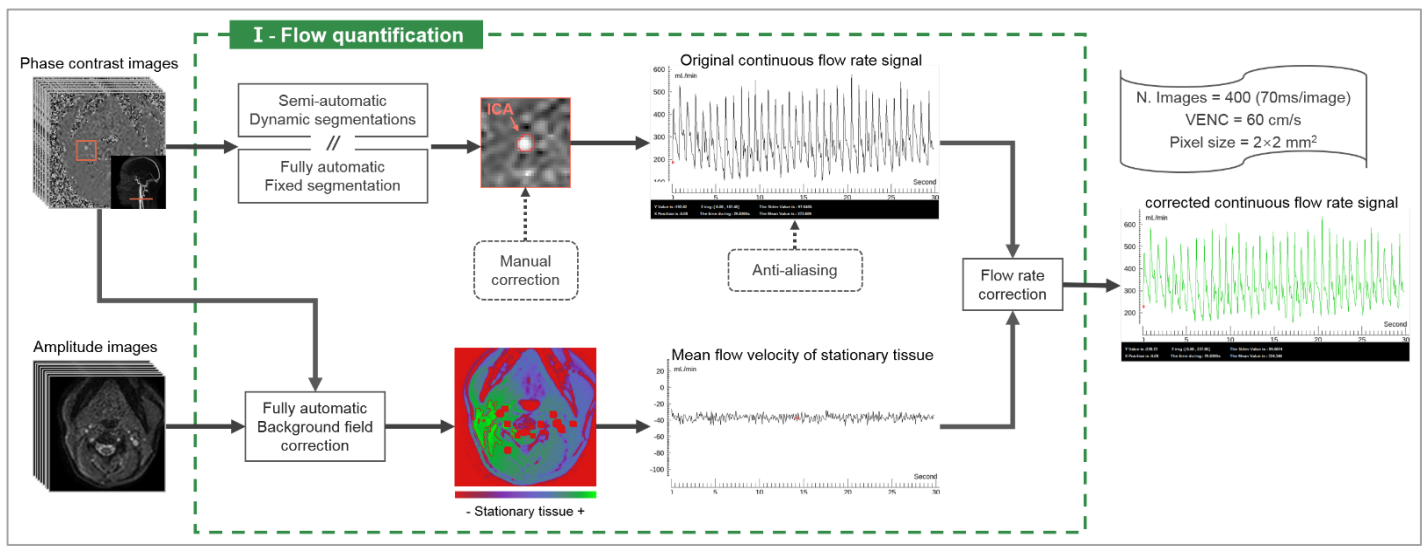

Figure 2: **Flow quantification workflow diagram.** 400 phase images and 400 amplitude images of the extracranial section were used as input parameters, and the corrected continuous flow rate signal of internal carotid artery (ICA) was used as the output result. Flow 2.0 can extract multiple signals from the ROI, and only the flow rate signal is shown in this figure. The stationary area around the ROI (red contour line) selected by the automatic background field correction algorithm.

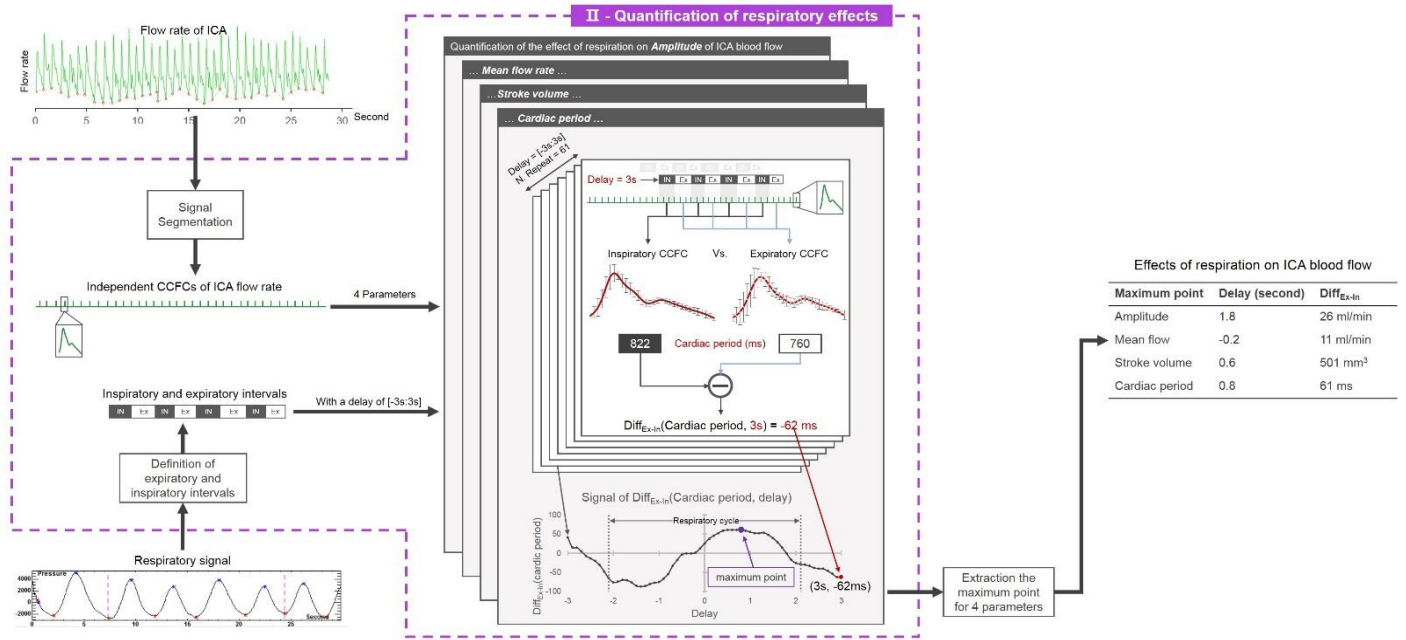

Figure 3: **Quantification of respiratory effects workflow diagram.** Flow rate signal of ICA and respiratory signal were used as input parameters. A $Diff_{Ex-In}$ *(parameter, delay)* signal is generated for each parameter (amplitude, mean flow, stroke volume, cardiac period) and its maximum value was taken to represent the intensity and delay of the effect of respiration on this parameter.

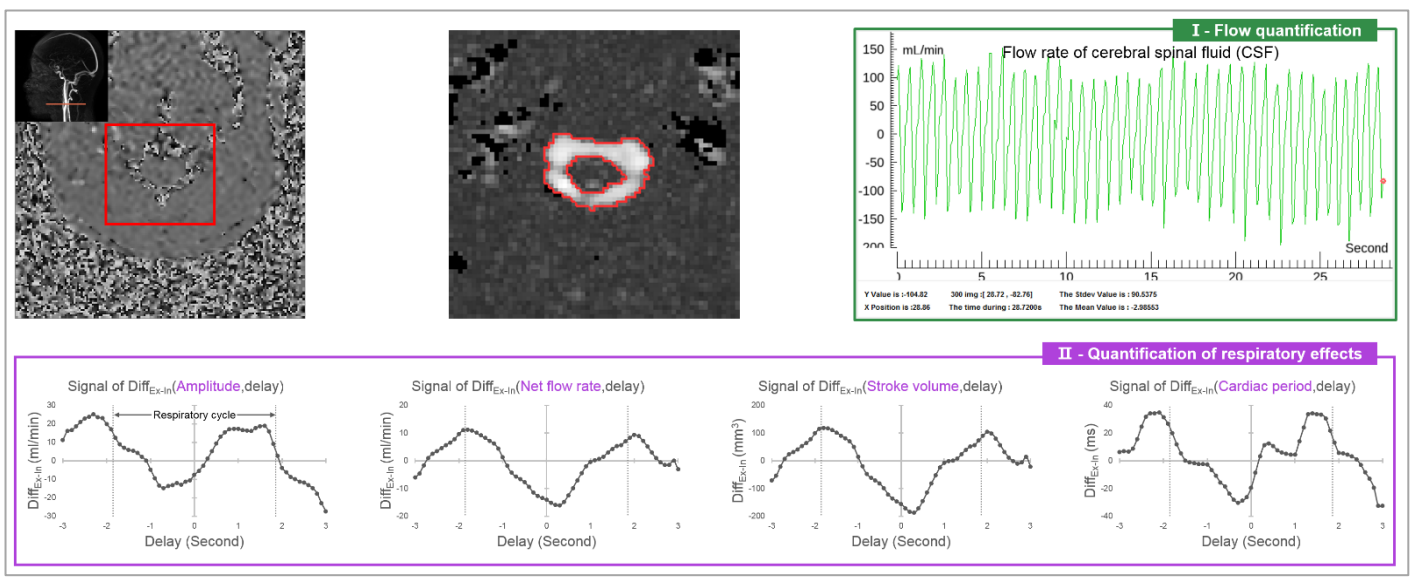

Figure 4: **An in vivo application of Flow 2.0 to quantify the effect of respiration on cerebrospinal fluid (CSF) in the extracranial section.** VENC = 5cm/s, imaging speed = 96ms/image, Number image = 300. The ROI of CSF can be accurately defined by a fully automated fixed segmentation algorithm, flow rate signal obtained after background field correction (green box). Combining the respiratory signal, the effect of respiration on the four parameters of CSF flow can be obtained (purple box).

Digital poster 2772

## References

1. Chen L, Beckett A, Verma A, & Feinberg D A. (2015). Dynamics of respiratory and cardiac CSF motion revealed with real-time simultaneous multi-slice EPI velocity phase contrast imaging. Neuroimage, 122, 281-287.
2. Aktas G, Kollmeier J M, Joseph A A, Merboldt K D, Ludwig H C, Gärtner J, ... & Dreha-Kulaczewski S (2019). Spinal CSF flow in response to forced thoracic and abdominal respiration. Fluids and Barriers of the CNS, 16(1), 1-8.
3. Balédent O, LIU P, Lokossou A, Sidy F, Metanbou S, Makki M (2019). Real-time phase contrast magnetic resonance imaging for assessment of cerebral hemodynamics during breathing. [Oral]. ISMRM 27th Annual Meeting & Exhibition in Montreal, Canada.
4. Lloyd R A, Butler J E, Gandevia S C, Ball I K, Toson B, Stoodley M A, & Bilston L E (2020). Respiratory cerebrospinal fluid flow is driven by the thoracic and lumbar spinal pressures. The Journal of Physiology, 598(24), 5789-5805.
5. Balédent O, & Idy-peretti I (2001). Cerebrospinal fluid dynamics and relation with blood flow: a magnetic resonance study with semiautomated cerebrospinal fluid segmentation. Investigative radiology, 36(7), 368-377.
6. Holland B J, Printz B F, & Lai W W (2010). Baseline correction of phase-contrast images in congenital cardiovascular magnetic resonance. Journal of cardiovascular magnetic resonance, 12(1), 1-7.

## Acknowledgements

This research was supported by equipex FIGURES (Facing Faces Institute Guilding Research), European Union Interreg REVERT Project, Hanuman ANR-18-CE45-0014 and Region Haut de France.

Thanks to the staff members at the Facing Faces Institute (Amiens, France) for technical assistance.

Thanks to David Chechin from Phillips industrie for his scientific support.